\documentclass[preprint,superscriptaddress,aps]{revtex4}
\usepackage{amsfonts,amsmath,amssymb}
\usepackage{amsfonts}
\usepackage{graphics}
\usepackage{graphicx}
\usepackage{epsf}

\newcommand{\bea}{\begin{eqnarray}}
\newcommand{\eea}{\end{eqnarray}}

\def\As{A\!\!\!/}
\def\ks{k\!\!\!/}

\def\ps{p\!\!\!/}

\def\ds{\partial\!\!\!/}
\begin{document}

\title{Vortex solutions in nonpolynomial scalar QED}

\author{F. C. E. Lima\footnote{E-mail: cleiton.estevao@fisica.ufc.br}}
\affiliation{Universidade Federal do Cear\'{a} (UFC), Departamento do F\'{i}­sica - Campus do Pici, Fortaleza, CE, C. P. 6030, 60455-760, Brazil.}

\author{A. Yu. Petrov\footnote{E-mail: petrov@fisica.ufpb.br}}
\affiliation{Universidade Federal da Para\'{i}ba (UFPB), Departamento de F\'{i}sica, Jo\~{a}o Pessoa, Para\'{i}ba, C. P. 5008, 58051-970, Brazil.}

\author{C. A. S. Almeida\footnote{E-mail: carlos@fisica.ufc.br}}
\affiliation{Universidade Federal do Cear\'{a} (UFC), Departamento do F\'{i}­sica - Campus do Pici, Fortaleza, CE, C. P. 6030, 60455-760, Brazil.}

\begin{abstract}
In order to investigate possible topological vortex structures in generalized models, we developed a perturbative generation approach for scalar-vector theories. We demonstrate explicitly that the dielectric permeability functions must have a nonpolynomial shape, i. e., the form of the logarithmic function. Basing on this result, we built models in $(2+1)D$ with logarithmic dielectric permeability in order to investigate the presence of topological vortex structures in a Maxwell model. This type of scalar-vector models is important because they can generate stationary field solutions in theories describing the dynamics of the scalar field. As examples, we chose models of the complex scalar field coupled to the Maxwell field. Subsequently, we investigated the model's Bogomol'nyi equations to describe the field configurations. Then, we demonstrate numerically, for an ansatz with rotational symmetry, that the solutions of the complex scalar field generating minimum energy configurations are topological structures depending on the parameters obtained in the perturbative generation of the vector-scalar theory.
\end{abstract}
\maketitle

\section{Introduction}

Topological vortex studies have attracted the attention of many researchers, see f.e. \cite{Vachaspati,Jackiw,Edeltein,Kim,Ghosh}, due to the possibility of their application in condensed matter physics \cite{Singh}. From the qualitative viewpoint, such topological structures are formed during a phase transition related to the breaking of some symmetry \cite{Giamarchi}. In this way, the BPS topological vortices arising within field theory context are similar to the known Abrikosov vortices known as characteristic phenomena in condensed matter physics \cite{Abrik,Abrik1}. Studies of these topological structures are based on some approaches. One of them is the known Bogomol'nyi-Prasad-Sommerfield (BPS) method \cite{Bogomol,PS}. Since the seminal paper by Bogomol'nyi \cite{Bogomol} where an innovative way for the study of the stability of classical field solutions was discussed, the study of topological defects has attracted the attention of many researchers, see f.e. \cite{Atmaja,Casana,Lima1,Lima2,Lima3}.

The motivations for studies of topological defects are extensive, since they can describe the dynamics of particles sets or even cosmological objects, as it was supposed in the seminal paper by Nielsen and Olesen \cite{Nilsen}. Among other consequences, it means that the analogue condensed matter models involving vortices can be applied for cosmological studies. In recent years, new models for topological defects  have emerged and been widely discussed in numerous papers, see f.e. \cite{Lee,Babichev,Adam,Atmaja1}. In general, these new generalized models are characterized by non-canonical kinetic terms, or in particular cases,  the gauge field is described by a non-canonical Lagrangian. It is important to mention here that, after the spontaneous symmetry breaking these nonstandard terms contribute also to the potential term necessary to produce the vortex.

In general, the vortex solutions in various generalized models display the similar topological behavior. In certain models, such structures possibly alter some physical quantities, for example, energy and magnetic field. It is interesting to mention that vortices arise within the study of the dynamics of scalar field or static vector field coupled to a gauge field in three-dimensional space-time. These structures find applications in several areas of physics \cite{Peterson,Sadler,Tong,Hu}, among them one can note detection of vortices in a lattice of linearly increasing potential \cite{Creutz}, study of configurations of magnetic flux tubes \cite{Haymaker}, etc. To justify presence of vortices, one can note as well that the vortex is a structure derived from dual superconductivity \cite{Haymaker}. This way, the vortices are related to the $U(1)$ symmetry of the model.

As we have mentioned already, the motivations for studying generalized models are vast. In this sense, wee note that the generalized models arise in the cosmological scenario within development of models for the inflationary expansion \cite{C}. In addition, these models are applied in order to solve the problem of cosmic coincidence \cite{C1}. It is interesting to note that in this scenario, generalized models can significantly change the dynamics of physical fields.

In our work, we are interested in studying the topological vortices of a model similar to the generalized Maxwell-Higgs model. However, in most of studies, non-polynomial scalar-vector models are introduced from the very beginning without any justification. In contrast to this, in order to propose a fundamental electrical permeability function, we built a model in $(3+1)D$ and developed a perturbative approach to scalar-vector fields theories. As a result, we observed that a possible form for the generalization function or the electrical permeability function is the logarithmic one. Therefore, motivated by this result and, by studies presented in ref. \cite{Edelstein1}, where it is shown that for an appropriate choice of the metric, we can reduce the BPS equations in $(3+1)D$ space-time to equations describing vortex-like structures, hence, we construct a new class of generalized nonpolynomial models. Thus, we propose a form for the electrical permeability function and show that the generalized models constructed with use of this function in the three-dimensional case admit topological solutions with interesting physical properties.

This work is organized as follows: in the section II, using the perturbative approach in scalar-vector theories, we find and show a form for the electrical permeability function. Then, in section III, we consider models in $(2+1)D$ generalized by the electrical permeability functions found in the previous section to study the topological vortices of the model through the BPS approach. In Section IV, we present some results and present some conclusions obtained throughout the work.

\section{Possible schemes for generating nonpolynomial scalar-vector models}

We start our study with proposing a scheme allowing for explaining the arising of non-polynomial scalar-vector terms. Our starting point will be the theory describing a spinor field coupled to scalar and vector fields, so we employ the methodology of dynamical generating new terms which has been applied in various field theory models, especially within the Lorentz-breaking context where it was used to generate the paradigmatic Carroll-Field-Jackiw term \cite{CFJ} (for a review on perturbative generation of new terms in Lorentz-breaking context see also \cite{ourLV}). We choose the action of the theory as
\bea
S=\int d^4x \bar{\psi}(i\ds-e\As- m(\phi))\psi+S_{scal}[\phi]+S_{Maxw}[A],
\eea
where $m(\phi)$ is a field-dependent mass of the spinor given by a some function of the scalar field $\phi$. This function is assumed to be arbitrary. We abandon the restriction of renormalizability of the theory since we intend to consider one-loop corrections (fermionic determinant) only, thus, the only divergences will arise in gauge and scalar sectors only. Moreover, we will be interested principally in terms like $F_{\mu\nu}F^{\mu\nu}f(\phi)$ which will provide the interesting exact solutions. Here $S_{scal}$ is a free action of the scalar field, and $S_{Maxw}$ is the Maxwell action of the electromagnetic field.  We assume the scalar $\phi$ to be slowly varying in the space-time, i.e. $\partial_{\mu}\phi$ is assumed to be small in comparison with any given energy scale, so, the derivative-dependent contributions in the scalar sector can be disregarded.

\begin{figure}[h]
    \centering
    \includegraphics[scale=1.1]{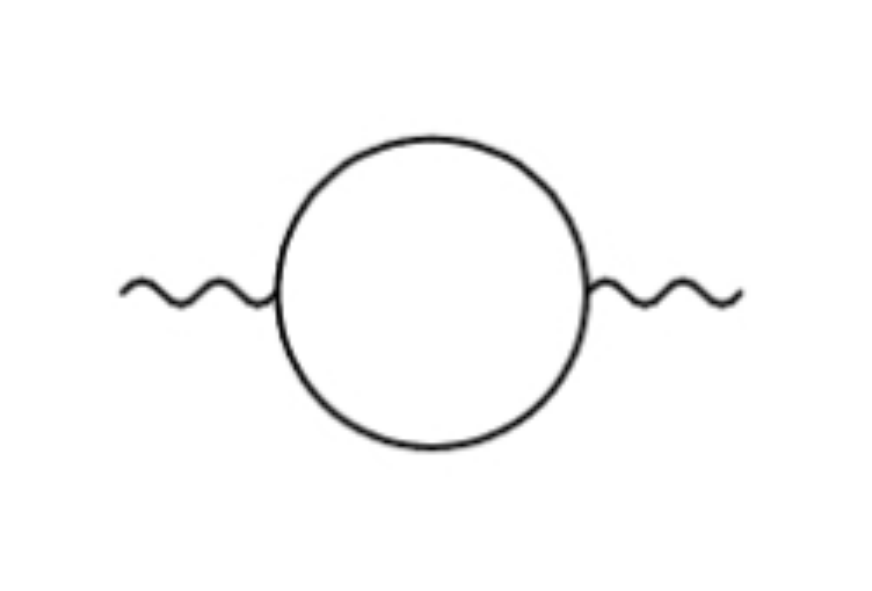}
    \caption{Feynman diagram.}
    \label{diagram}
\end{figure}

To generate the term $F_{\mu\nu}F^{\mu\nu}f(\phi)$, we start with the two-point function of the gauge field. It is given by the Feynman diagram depicted at Fig. \ref{diagram}.

Its contribution is given by
\bea
\label{defga}
\Gamma_2=-\frac{e^2}{2}\int\frac{d^4p}{(2\pi)^4}A_{\mu}(-p)\Pi^{\mu\nu}(p)A_{\nu}(p),
\eea
where the minus sign is caused by the fact that the loop is fermionic, $1/2$ is a combinatoric coefficient, and the self-energy tensor is
\bea
\Pi^{\mu\nu}={\rm tr}\int\frac{d^4k}{(2\pi)^4}\gamma^{\mu}\frac{1}{\ks-m(\phi)}\gamma^{\nu}\frac{1}{\ks+\ps-m(\phi)}.
\eea
We can perform the calculations in a whole analogy with \cite{Fuji}, with the only difference is that the mass is a nontrivial function of the scalar fields. So, let us calculate the $\Pi^{\mu\nu}$. Proceeding as in \cite{Fuji}, we extract some common multipliers and simplify the expression:
\bea
\label{2point}
\Pi^{\mu\nu}&=&4i\frac{\mu^{4-d}}{(4\pi)^{d/2}}\Gamma(2-\frac{d}{2})\Big[\int_0^1dx [m^2(\phi)-p^2x(1-x)]^{d/2-2}\times\nonumber\\&\times&
\left[-2p^{\mu}p^{\nu}x(1-x)+\eta^{\mu\nu}p^2x(1-x)+\eta^{\mu\nu}m^2(\phi)\right]-\nonumber\\
&-&\eta^{\mu\nu}
\int_0^1dx [m^2(\phi)-p^2x(1-x)]^{d/2-1}\Big].
\eea

We choose $d=4-\epsilon$ as above, so, $2-\frac{d}{2}=\frac{\epsilon}{2}$. First, we concentrate on the divergent part, i.e. everywhere except of the gamma function we put $d=4$. After the inverse Wick rotation (multiplying by $-i$) we find that the divergent part is $\phi$ independent since the divergent terms depending on $m^2(\phi)$ cancel out:
\bea
\label{2pointqed}
\Pi^{\mu\nu}&=&\frac{8}{(4\pi)^2}\Gamma(\frac{\epsilon}{2})\int_0^1dx  x(1-x)
\left[-p^{\mu}p^{\nu}+\eta^{\mu\nu}p^2\right]=\frac{1}{6\pi^2\epsilon}\left[-p^{\mu}p^{\nu}+\eta^{\mu\nu}p^2\right].
\eea

Repeating the argumentation of \cite{Fuji}, we see that the divergent part of the one-loop correction can be presented as $\frac{e^2}{6\pi^2\epsilon}{\cal L}_{Maxw}$ which is the paradigmatic result of QED which clearly can be cancelled by the simple wave function renormalization.

Now, let us consider the finite part of (\ref{2point}) which is necessary for our purposes. To do it, we take into account only terms which neither vanish nor diverge at $\epsilon\to 0$. It implies the following contribution to the effective action:
\bea
\Gamma^{(1)}=-\frac{e^2}{12\pi^2}\int d^4x F_{\mu\nu}F^{\mu\nu}(\ln\frac{m(\phi)^{2}}{\mu^2}+C),
\eea
where $C$ is a some field-independent number which can be absorbed into rescaling of $\mu$ or into {\bf a} finite renormalization of the kinetic term.

So, the net result of our calculation is the arising of the one-loop contribution
\bea
\label{netres}
\Sigma=-\frac{a}{4}F_{\mu\nu}F^{\mu\nu}\ln\frac{m(\phi)^{2}}{\mu^2},
\eea
with $a$ is a some number.
Therefore, now we can make various assumptions for the $m(\phi)^2$ to obtain desired scalar-vector actions. The most natural example is $m(\phi)=m_0+h\phi$, or simply $m(\phi)=h\phi$. The last choice is more advantageous since it yields the $F^2\ln\phi$ logarithmic term.

One more possibility to obtain new scalar-vector terms is as follows: we consider the action
\bea
S=\int d^4x \bar{\psi}(i\ds-ef(\phi\phi^*)\As- m(\phi^*\phi))\psi+S_{scal}[\phi]+S_{Maxw},
\eea
i.e., we have together with the field-dependent mass $m(\phi^*\phi)$, the field-dependent coupling $ef(\phi\phi^*)$ which can be related with dielectric (magnetic) permeability, see discussion in \cite{dual} and references therein. In this case, after cancellation of the one-loop divergence with the appropriate counterterms, we have the finite contribution
\bea
\label{fincont}
\Gamma^{(1)}=-\frac{e^2}{12\pi^2}\int d^4x f^2(\phi^*\phi) F_{\mu\nu}F^{\mu\nu}\ln\frac{m^2(\phi^*\phi)}{\mu^2},
\eea
where again we absorbed the finite constant into redefinition of $\mu$. We immediately see that the Lagrangian obtained in \cite{Lima1} looking like
\bea
\label{previous}
{\cal L}_0=c|\phi|^2\ln\frac{|\phi|^2}{\mu^2}F_{\mu\nu}F^{\mu\nu},
\eea
is replayed if $f^2(\phi\phi^*)\propto \phi\phi^*$, and $m^2(\phi\phi^*)\propto\phi\phi^*$.
Therefore we have various possibilities allowing to replay the theory (\ref{previous}) through making an appropriate choice for $m(\phi\phi^*)$ and $f(\phi\phi^*)$ which can be done in order to match experimental results.

First, we can assume that our scalar field is real, $\phi=\phi^*$. In this case we can choose $m(\phi)=h\phi$, and $f(\phi)=\lambda\phi$. However, while in this case we can have non-polynomial (logarithmic) terms in the effective action, including logarithms of more sophisticated functions, f.e. polynomial ones, this case is reproduced only within certain approximations, see next sections.

Second, we can choose $m(\phi\phi^*)=h|\phi|$, and $f(\phi\phi^*)=\lambda|\phi|$ (where $|\phi|=\sqrt{\phi\phi^*}$). In this case we reproduce exactly the desired term (\ref{previous}). However, such a choice needs some qualitative justification and seems to be a construction introduced {\it ad hoc}. A some explanation will be presented in next sections. Alternatively, the logarithmic term $\ln\frac{\phi\phi^*}{\mu^2}$ can arise if we choose $m(\phi\phi^*)=\frac{\phi\phi^*}{a}$, where $a$ is an inverse Yukawa coupling, so, our vertex is $\frac{1}{a}\bar{\psi}\phi\phi^*\psi$, and $a$ is afterwards reabsorbed into redefinition of $\mu$. 

Third, the $|\phi|^2$ factor can be generated in (\ref{fincont}) also as follows: together with $m(\phi\phi^*)=\frac{\phi\phi^*}{a}$, we assume $f(\phi\phi^*)=1+\frac{\phi\phi^*}{2\Lambda^2}$, with $\Lambda\to\infty$. In this case (\ref{fincont}) takes the form
\bea
\label{fincont1}
\Gamma^{(1)}=-\frac{e^2}{24\pi^2}\int d^4x  (1+\frac{\phi\phi^*}{\Lambda^2})F_{\mu\nu}F^{\mu\nu}\ln\frac{\phi^*\phi}{a\mu}+O(\Lambda^{-4}),
\eea
which includes a sum of the desired term (\ref{previous}) and the simple term $F^2\ln|\phi|$ (\ref{netres}). In this case we avoid rather unnatural square root functions. So, now our main task consists in obtaining exact solutions for these theories and their comparison with results of \cite{Lima1}.

\section{Generalized vortex structure}

Now, we start with explicit study of vortex-like solutions for various generalizations of electrodynamics with non-trivial electric permeability. We consider two cases: permeability $\epsilon_1=\rho\ln(m|\phi|^{2}/\mu^{2})$ and permeability $\epsilon_2=\rho f(|\phi|^2)\ln(m|\phi|^{2}/\mu^{2})$.

\subsection{Electric permeability of the type $\rho\ln(m(|\phi|)^{2}/\mu^{2})$}

Let us consider the first example. Motivated by the discussion presented above of QED we will study models in $(2+1)D$, described by the action \cite{Lima1}:
\begin{equation}
\label{action1}
    S=\int\, d^{3}x \bigg[-\frac{\rho}{4}\ln\bigg(\frac{m(\phi)^{2}}{\mu^{2}}\bigg)F_{\mu\nu}F^{\mu\nu}+|D_{\mu}\phi|^{2}-\mathcal{V}\bigg].
\end{equation}
where $\rho$ is a dimensionless adjustable parameter. We note that this kind of models can be obtained on the base of the one-loop result (\ref{netres}) under a reduction of the four-dimensional theory to the three-dimensional one in a manner similar to  \cite{CasFerRed}. To do it we can ``freeze'' the extra spatial coordinate, say $z$, so that all fields display no dependence on $z$ more, and the axial gauge $A_3=0$ is imposed. Besides of this, the action of this form can be treated as a certain reminiscence of the scalarization mechanism used within gravity context \cite{Sotir} where new terms equal to gravity topological invariants multiplied by some functions of a scalar field arise.

The generalized models was introduced for the first time by Lee and Nam where they discuss the soliton solutions of an Abelian Higgs model \cite{LeeNam}. Further, many other studies of models with generalized dynamics were performed. In general, it is interesting to generalize a model of a complex scalar field coupled to a gauge field, in order to generate new classical solutions. Typically, new models are constructed through multiplicative terms representing themselves as functions of the fields. In some papers, see f.e. \cite{dual} and references therein, the function multiplying the Maxwell term is treated as a generalizations of electric permeability. It is natural to assume this function to be positively definite one.

Other studies have already discussed the vortex structures in theories with logarithmic interaction arising due to spontaneous breaking symmetry. For example, the paper \cite{Lima1} opens the door to new possibilities for studies of more general models with logarithmic electrical permeability.

In this paper, being motivated by the perturbative generation of scalar-vector terms (see the previous section), we present topological structures for some forms of vortices solutions with a nonpolynomial electrical permeability function. This theory will involve a nonlinear coupling of scalar fields to a gauge field so that one can consider various forms of $m(|\phi|)^{2}$, and for each form of this parameter the vortices admit different physical properties.

Let us make our definitions. The field $\phi$ is the complex scalar field, so $|\phi|^{2}=\bar{\phi}\phi$. Here we use an overline to denote the complex conjugation of the field. The Abelian gauge field is denoted by $A_{\mu}$. The electromagnetic tensor is described by $F_{\mu\nu}$, where it is defined
\begin{align}
    F_{\mu\nu}=\partial_{\mu}A_{\nu}-\partial_{\nu}A_{\mu}.
\end{align}
We also define the covariant derivative $D_{\mu}=\partial_{\mu}+ieA_{\mu}$. We will assume the metric signature  $g_{\mu\nu}=$ diag$(+, -, -)$ and use the natural system of units where $\hbar=c=1$.

It is interesting to mention that in the action (\ref{action1}) we assume that $m(\phi)^{2}=m(|\phi|^{2})$. This equality can be used since that when we study topological structures in $(2+1)D$, we assume the ansatz
\begin{align}
\label{ansatz1}
    \phi=g(r)\text{e}^{in\theta},
\end{align}
so, in the limit $n\theta\ll 1$, we will have $\bar{\phi}=\phi$. Thus, for this choice of the complex scalar field, we can without loss of generality, write the action according to the expression (\ref{action1}).

From now on, we will look for solutions for the gauge field described by an ansatz of the form:
\begin{align}
\label{ansatz2}
    \Vec{A}=-\frac{1}{er}[a(r)-n]\hat{\text{e}}_{\theta}.
\end{align}

We started our study of the BPS vortices, considering the generalized Lagrangian density as:
\begin{align}
    \mathcal{L}=-\frac{\rho}{4}\ln\bigg(\frac{m(|\phi|)^{2}}{\mu^{2}}\bigg)F_{\mu\nu}F^{\mu\nu}+|D_{\mu}\phi|^{2}-\mathcal{V}(|\phi|).
\end{align}
where we remember that the parameter $\mu$ within the logarithm is a positively definite normalization parameter used to adjust the canonical dimension of the magnetic permeability term.

The equations of motion of the model are:
\begin{align}
\label{motion1}
    D_{\mu}D^{\mu}\phi+\frac{\phi}{2|\phi|}\bigg[\frac{\rho m_{|\phi|}}{2m(|\phi|)}F_{\mu\nu}F^{\mu\nu}+\mathcal{V}_{|\phi|}\bigg]=0,
\end{align}

\begin{align}
\label{motion2}
    J^{\nu}=\partial_{\mu}\bigg[\rho\ln\bigg(\frac{m(|\phi|)^{2}}{\mu^{2}}\bigg)F^{\mu\nu}\bigg],
\end{align}
where we define $m_{|\phi|}=dm/d|\phi|$  and $\mathcal{V}_{|\phi|}=\partial\mathcal{V} /\partial |\phi|$. If we have $\nu=0$ in eq. (\ref{motion2}) we get Gauss's law. Thus, we conclude that the vortices are electrically neutral.

The conserved current of the model is described by:
\begin{align}
\label{current2}
    J_{\mu}=ie(\overline{\phi}D_{\mu}\phi-\phi\overline{D_{\mu}\phi}).
\end{align}

Using the spacetime translation symmetry, we construct the energy-momentum tensor, namely,
\begin{align}
    T_{\mu\nu}=\rho\ln\bigg(\frac{m(|\phi|)^{2}}{\mu^{2}}\bigg)F_{\mu\lambda}F^{\lambda}\, _{\nu}+\overline{D_{\mu}\phi}D_{\nu}\phi+\overline{D_{\nu}\phi}D_{\mu}\phi-g_{\mu\nu}\mathcal{L}.
\end{align}

To study static and rotationally symmetric solutions, we consider the equations (\ref{ansatz1}) and (\ref{ansatz2}), for which we assume the boundary conditions:
\begin{align}\nonumber
\label{boundary}
    &g(0)=0, \hspace{2cm} g(\infty)=\xi,\\
    &a(0)=n, \hspace{2cm} a(\infty)=0.
\end{align}
where $\xi$ is the v.e.v., and the $n$ parameter is related to the model's vorticity.

Using the expression (\ref{ansatz2}), we find that the magnetic field of the vortex is
\begin{align}
\label{B}
    B=-F_{12}=-\frac{a'(r)}{er},
\end{align}
therefore, we observe that the vortex of the model has a magnetic flux given by
\begin{align}
    &\Phi_{B}=\int_{S}\int\Vec{B}\cdot\Vec{dS}=-\int_{0}^{2\pi}\int_{0}^{\infty}F^{12}rdrd\theta=-\frac{2\pi}{e}[a(\infty)-a(0)].
\end{align}

Using (\ref{boundary}) we conclude that the magnetic flux of the generalized vortex is
\begin{align}
    \Phi_{B}=\frac{2\pi n}{e},
\end{align}
therefore, the model flux will be quantized.

Now, our main objects of study will be the field variables, i. e., the radial functions $g(r)$ and $a(r)$. With this in mind, we use the equations (\ref{ansatz1}) and (\ref{ansatz2}) to rewrite the equation of motion of the model in terms of the field variables. In this way, we obtain that the equations of motion for the static fields will be given by:
\begin{align}
   \frac{1}{r}[rg(r)]'=\frac{\rho m_{g}a'(r)^{2}}{2m(g)e^{2}r^{2}}+\frac{a(r)^{2}g(r)}{r^{2}}+\mathcal{V}_{g}, 
\end{align}
and
\begin{align}
    r\bigg[\ln\bigg(\frac{m(g)}{\mu}\bigg)\bigg]=\frac{ea(r)g(r)^{2}}{\rho}.
\end{align}

We are interested in the study of static field configuration with finite energy. To do this, we turn our attention to the study of the energy density of these fields, namely,
\begin{align}
\label{energy1}
    \mathcal{E}=T_{00}=\frac{\rho a'(r)^{2}}{e^{2}r^{2}}\ln\bigg(\frac{m(g)}{\mu}\bigg)+g'(r)^{2}+\frac{a(r)^{2}g(r)^{2}}{r^{2}}+\mathcal{V}(g).
\end{align}

Seeking to satisfy the BPS limit, we reorganize the previous expression writing
\begin{align}\nonumber
\label{energy2}
    \mathcal{E}=&\frac{\rho}{2}\ln\bigg(\frac{m(g)^{2}}{\mu^{2}}\bigg)\bigg[\frac{a'(r)}{er}\pm\frac{e(\xi^{2}-g(r)^{2})}{\rho\ln(m(g)^{2}/\mu^{2})}\bigg]^{2}+\bigg(g'(r)\mp\frac{a(r)g(r)}{r}\bigg)^{2}+\mathcal{V}\\
    & -\frac{e^{2}}{2}\frac{(\xi^{2}-g(r)^{2})^{2}}{\rho\ln(m(g)^{2}/\mu^{2})}\mp\frac{1}{r}[a(r)(\xi^{2}-g(r)^{2})]'.
\end{align}

Here, we chose the potential in the form
\begin{align}
    \mathcal{V}=\frac{e^{2}}{2}\frac{(\xi^{2}-g(r)^{2})^{2}}{\rho\ln(m(g)^{2}/\mu^{2})},
\end{align}
then, the equation (\ref{energy2}) is simplified significantly.

Bearing in mind that the energy of the vortex is described by the integration of energy density over all space, we arrive at
\begin{align}\nonumber
\label{energy3}
    E=&2\pi\int_{0}^{\infty}\,\rho r\ln\bigg(\frac{m(g)}{\mu}\bigg)\bigg[\frac{a'(r)}{er}\pm\frac{e(\xi^{2}-g(r)^{2})}{\rho\ln(m(g)^{2}/\mu^{2})}\bigg]^{2}\, dr+2\pi\int_{0}^{\infty}\, r\bigg(g'(r)\mp\frac{a(r)g(r)}{r}\bigg)^{2}\, dr\\
    &+E_{BPS},
\end{align}
where $E_{BPS}$ is the Bogomol'nyi energy, defined by
\begin{align}
    E_{BPS}=\mp 2\pi\int_{0}^{\infty}\, [a(r)(\xi^{2}-g(r)^{2})]'\, dr=2\pi |n|\xi^{2}.
\end{align}

We note that the energy is limited from below by the limit $E_{BPS}$, i. e., $E\geq E_{BPS}$. At the boundary of saturation, $E=E_{BPS}$, we find that the static fields obey first-order equations:
\begin{align}
    g'(r)=\pm\frac{a(r)g(r)}{r},
\end{align}
and
\begin{align}
    \frac{a'(r)}{er}=\mp\frac{e(\xi^{2}-g(r)^{2})}{2\rho\ln(m(g)/\mu)}.
\end{align}

The above equations are known as the model's Bogomol'nyi equations. At the BPS limit, the energy density of the fields looks like
\begin{align}
    \mathcal{E}_{BPS}=\frac{2\rho a'(r)^{2}}{e^{2}r^{2}}\ln\bigg(\frac{m(g)}{\mu}\bigg)+2g'(r)^{2}.
\end{align}

\subsubsection{Case $m(|\phi|)=h|\phi|$}

The simplest case we can think of for studying topological vortices is to assume that the function $m(|\phi|)=h|\phi|$, with $h$ a constant to be defined later. Obviously, as we assumed at the beginning of the paper the contribution for $n\theta\ll 1$, we are interested in the investigation of the topological structures of lower vorticity, i. e., $n =1$. From this point on, we observe that in this case the BPS equations of the model are reduced to the
\begin{align}
\label{BPS3}
    g'(r)=\pm\frac{a(r)g(r)}{r},
\end{align}
and
\begin{align}
\label{BPS4}
    \frac{a'(r)}{er}=\mp\frac{e(\xi^{2}-g(r)^{2})}{2\rho\ln\bigg(\frac{hg(r)}{\mu}\bigg)}.
\end{align}

The BPS energy density is
\begin{align}
\label{EBPS2}
    \mathcal{E}_{BPS}=\frac{2\rho a'(r)^{2}}{e^{2}r^{2}}\ln\bigg(\frac{hg(r)}{\mu}\bigg)+2g'(r)^{2}.
\end{align}

So, we decoupled the equations describing the radial parts of the scalar field, $g(r)$, and of the gauge one $a(r)$. In this way, we write the equation for $g(r)$, namely,
\begin{align}
\label{VField2}
    g''(r)-\frac{g'(r)^{2}}{g(r)}+\frac{g'(r)}{r}+\frac{e^{2}(\xi^{2}-g(r)^{2})g(r)}{2\rho\ln\bigg(\frac{hg(r)}{\mu}\bigg)}=0.
\end{align}

\begin{figure}[ht!]
    \centering
    \includegraphics[scale=0.55]{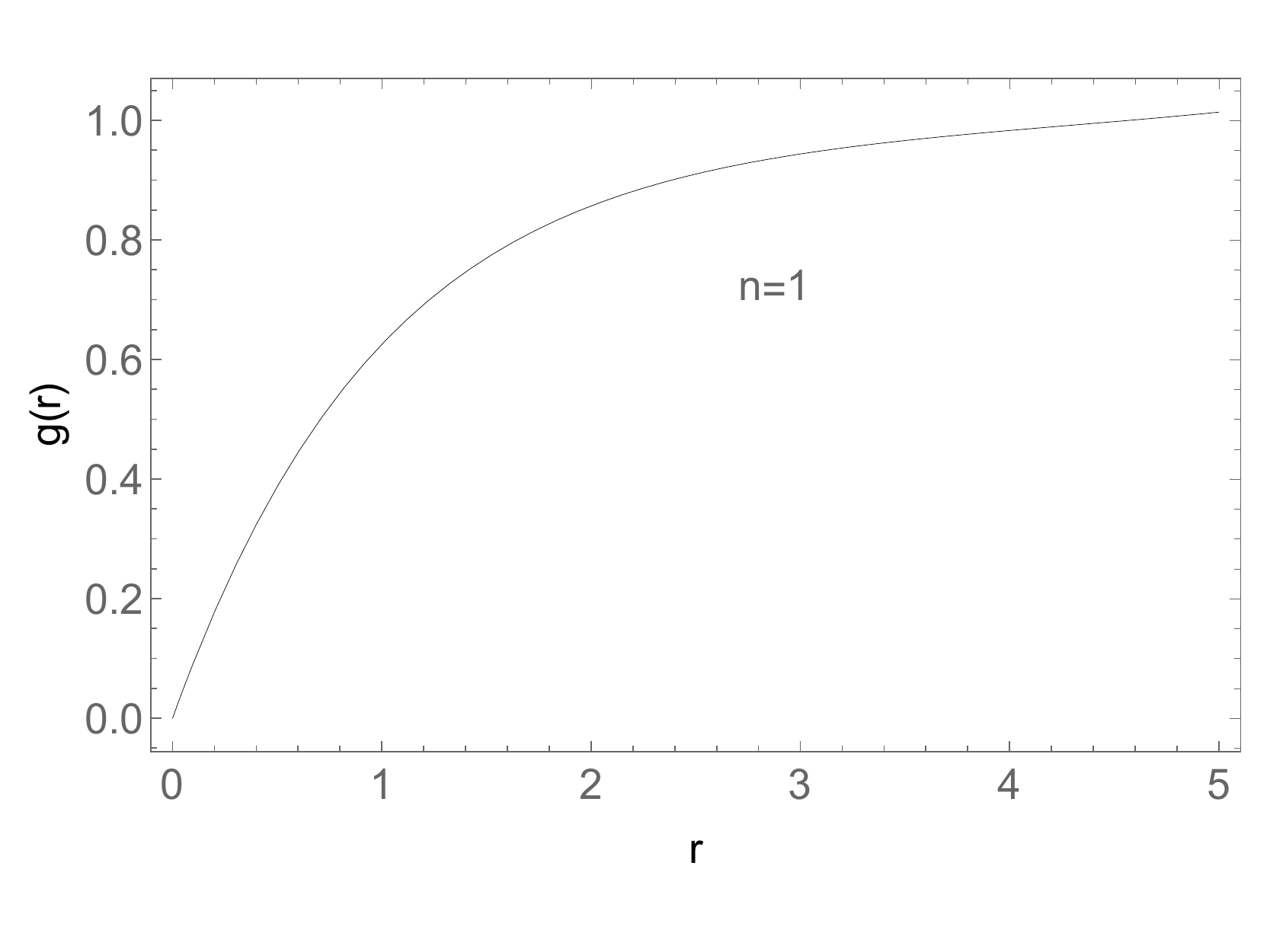}
    \caption{Behavior of the variable field corresponding to the complex scalar field when $e=\xi=1$, $\rho=10^{3}$, $h=10^{- 3}$ and $\mu=10^{-10}$.}
    \label{fig5}
\end{figure}

We consider the topological boundary for the $g(r)$ field, and must investigate the vortex structures admitted by the model.
\vspace{0.5cm}
\paragraph{Numerical results.}
From now on, we have focused on the numerical investigation of the topological vortices of the model. In order to achieve our goal, we perform numerical solution for the equation (\ref{VField2}). The corresponding result is shown in fig. \ref{fig5}.

\begin{figure}[ht!]
    \centering
    \includegraphics[scale=0.55]{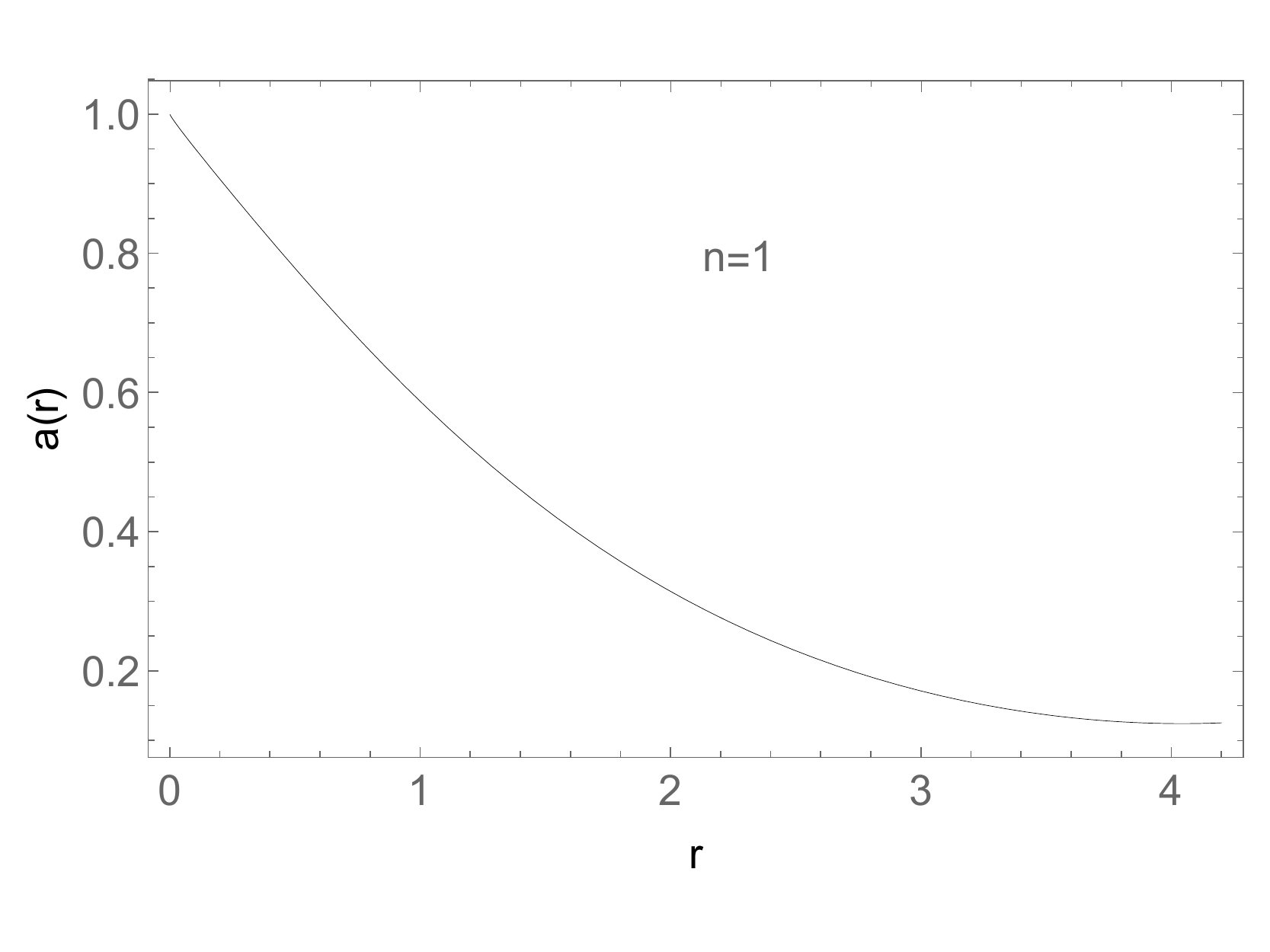}
    \caption{Solution of the gauge field when $e=\xi=1$, $\rho=10^{3}$, $h=10^{- 3}$ and $\mu=10^{-10}$.}
    \label{fig6}
\end{figure}

With the solution of the complex scalar field depicted at fig. \ref{fig5}, we can, with the use of eq. (\ref{BPS3}), investigate the solution for the gauge field numerically. The result for $a(r)$ is depicted in fig. \ref{fig6}.

\begin{figure}[ht!]
    \centering
    \includegraphics[scale=0.55]{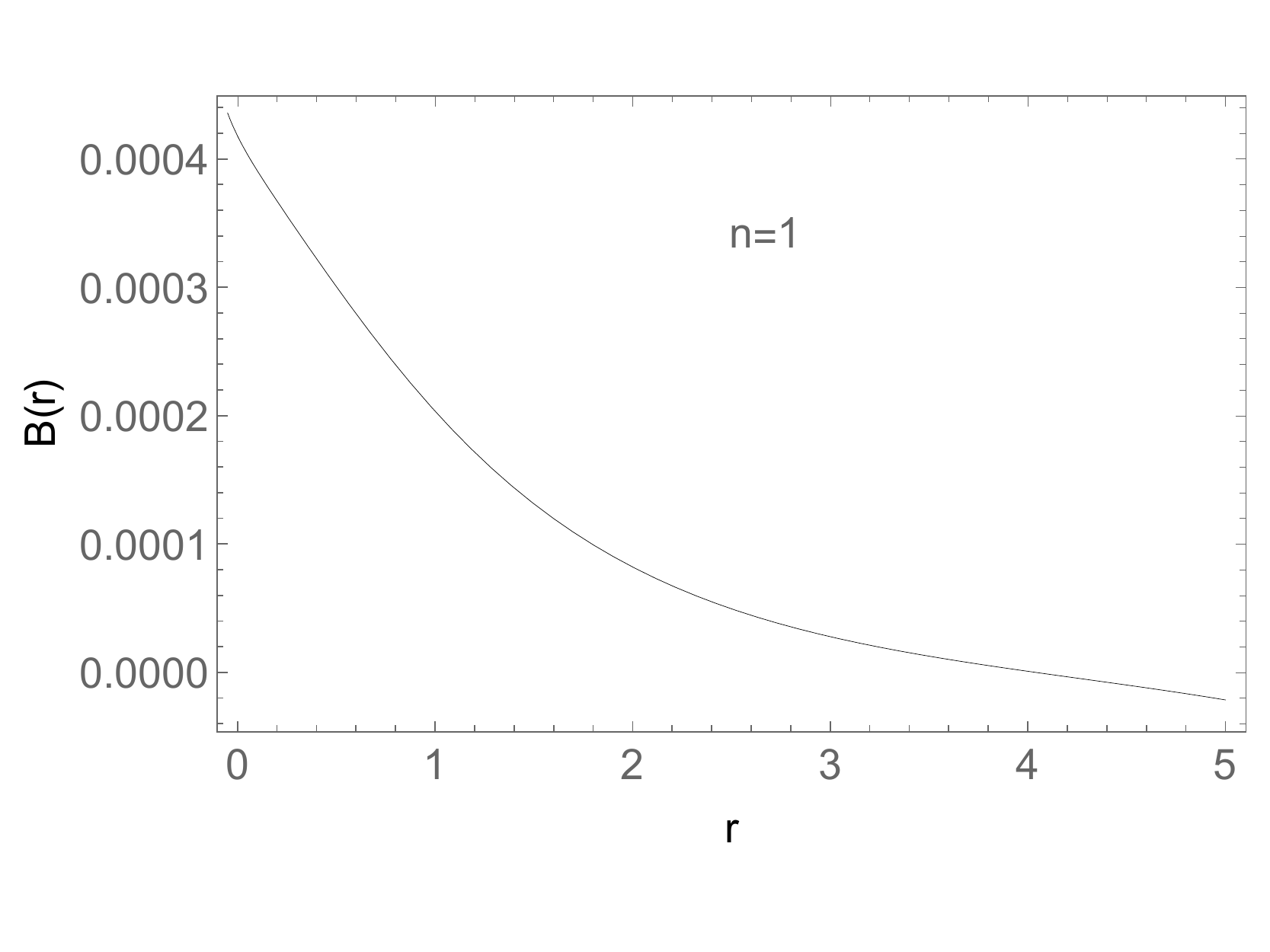}
    \caption{Magnetic fields associated with the BPS vortex.}
    \label{fig7}
\end{figure}

From the eq. (\ref{B}), we can find the numerical solution of the magnetic field responsible for the flux of the vortex. The magnetic field is shown in fig. \ref{fig7}.

\begin{figure}[ht!]
    \centering
    \includegraphics[scale=0.55]{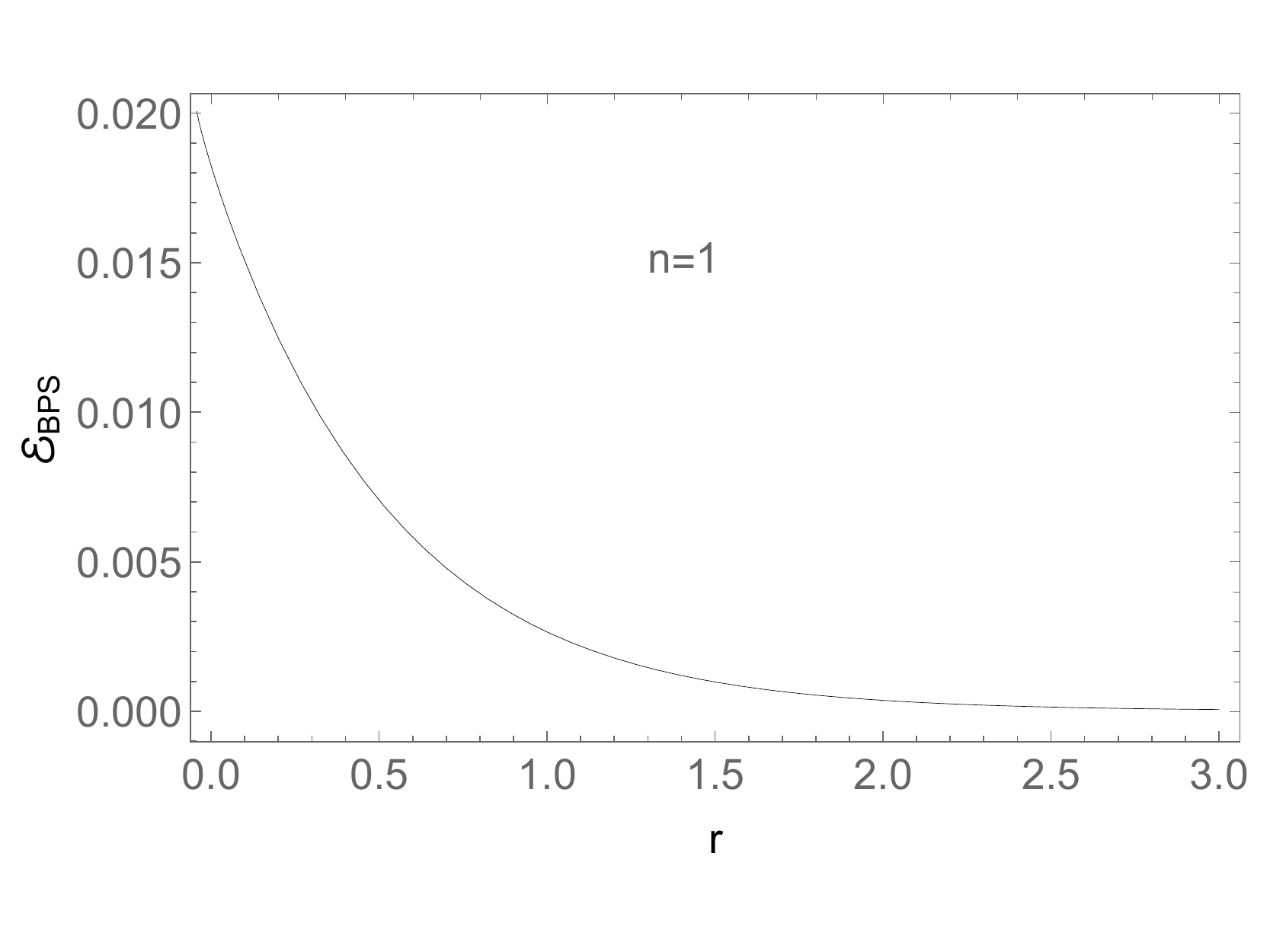}
    \caption{BPS energy density of the vortex.}
    \label{fig8}
\end{figure}

The model's BPS energy, can be found from eq. (\ref{EBPS2}). The corresponding numerical solution shown in fig. \ref{fig8}.

From the numerical results for vortex structures, we observed that these structures have a more intense magnetic flux near the center $r=0$. This fact ends up making the vortex more energetic in the system's origin. We observed that the magnetic field produced by the vortex has a behavior similar to the BPS energy density, differing significantly in its magnitude. We observed that the BPS vortices, despite having a ``weak magnetic field'', possess a significant energy. This leads us to the hypothesis that this happens due to the electrical permeability displaying a logarithmic behavior.

\subsubsection{The case: $m(|\phi|)=m_{0}+h|\phi|$}

In this case, we see that the model's BPS equations are now written in the form:
\begin{align}
\label{BPS1}
    g'(r)=\pm\frac{a(r)g(r)}{r},
\end{align}
and
\begin{align}
\label{BPS2}
    \frac{a'(r)}{er}=\mp\frac{e(\xi^{2}-g(r)^{2})}{2\rho\ln\bigg(\frac{m_{0}+hg(r)}{\mu}\bigg)},
\end{align}
with $m_{0}$ and $h$ being constants to be defined.

In this case, the BPS energy density of the vortices will be described by:
\begin{align}
\label{EBPS}
    \mathcal{E}_{BPS}=\frac{2\rho a'(r)^{2}}{e^{2}r^{2}}\ln\bigg(\frac{m_{0}+hg(r)}{\mu}\bigg)+2g'(r)^{2}.
\end{align}

To describe the topological solutions we investigate the solutions of the equations (\ref{BPS1}) and (\ref{BPS2}). For this, we start by decoupling the BPS equations. Thus, we obtain that the equation describing the complex scalar field is
\begin{align}
\label{VField}
    g''(r)-\frac{g'(r)^{2}}{g(r)}+\frac{g'(r)}{r}+\frac{e^{2}(\xi^{2}-g(r)^{2})g(r)}{2\rho\ln\bigg(\frac{m_{0}+hg(r)}{\mu}\bigg)}=0.
\end{align}

The behavior of the $g(r)$ must respect the boundary conditions
\begin{align}
\label{conditionG}
    g(0)=0 \hspace{1cm} \text{and} \hspace{1cm} g(\infty)=\xi.
\end{align}

\begin{figure}[ht!]
    \centering
    \includegraphics[scale=0.55]{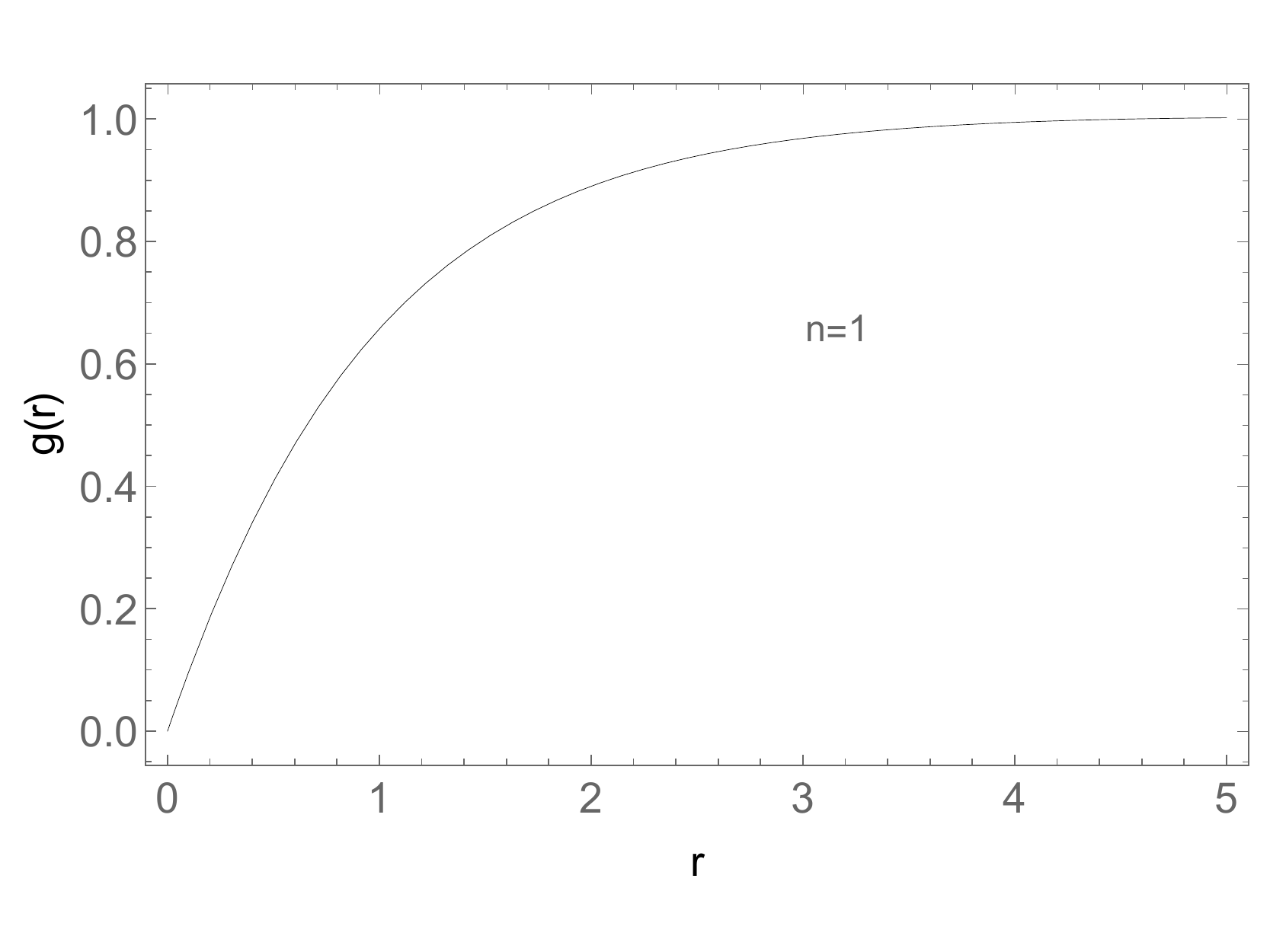}
    \caption{Behavior of the variable field corresponding to the complex scalar field when $e=m_{0}=\xi=1$, $\rho=10^{3}$, $h=10^{- 3}$ and $\mu=10^{-10}$.}
    \label{fig1}
\end{figure}

\paragraph{Numerical results.}
The equation (\ref{VField}) can be considered only with use of a numerical analysis. To do this, we use the interpolation method. Having this in mind, we turn our attention to studying the topological vortex structures of the model. Thus, we start with considering eq. (\ref{conditionG}), and assume the lower value for the vorticity, i. e., $n=1$. In this way, we obtain the vortex solution presented in fig. \ref{fig1}.

\begin{figure}[ht!]
    \centering
    \includegraphics[scale=0.55]{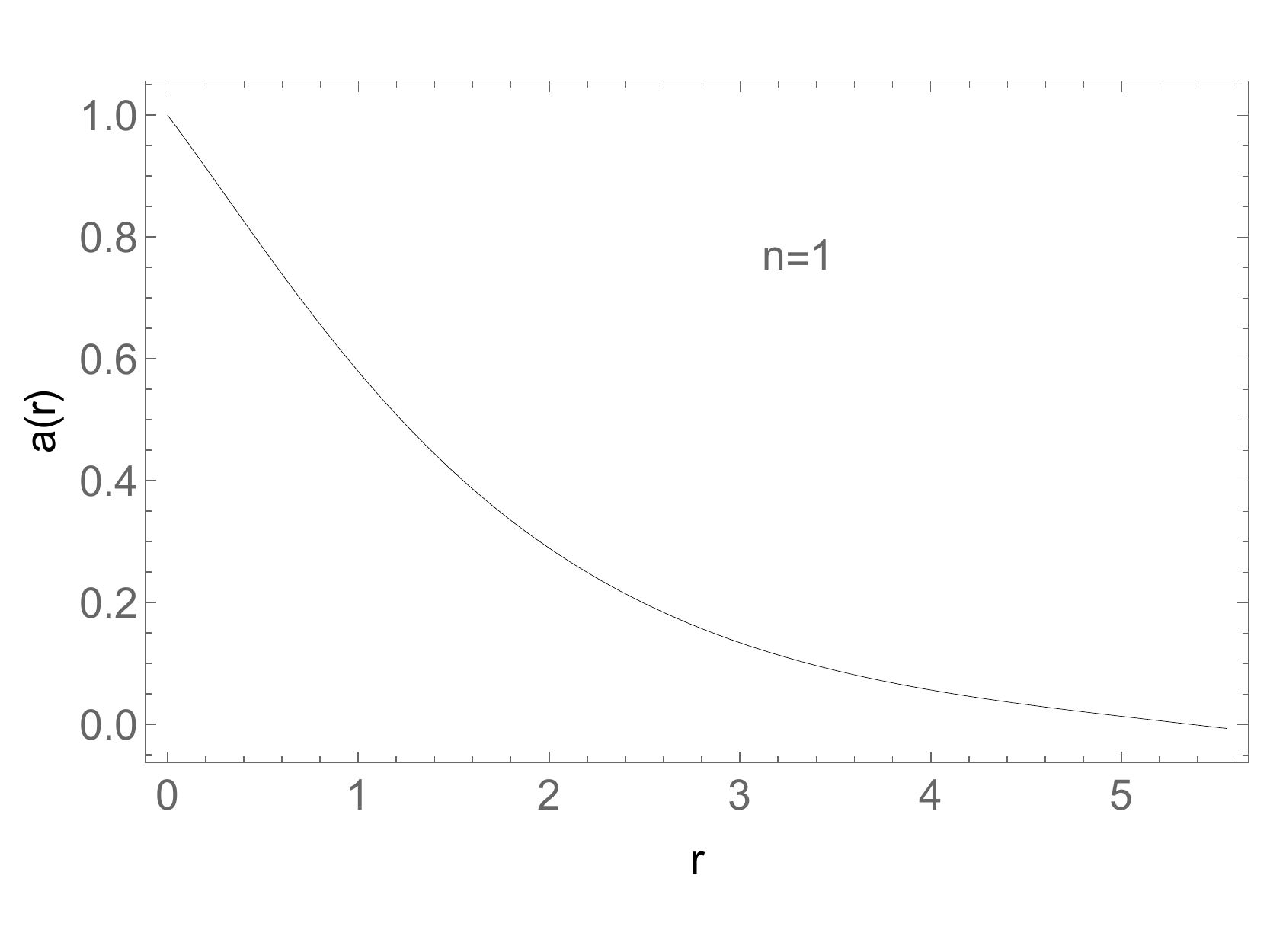}
    \caption{Behavior of the gauge field when $e=m_{0}=\xi=1$, $\rho=10^{3}$ $h=10^{- 3}$ and $\mu=10^{-10}$.}
    \label{fig2}
\end{figure}

Once we find the solution of the complex scalar field (fig. \ref{fig1}), we can use the BPS equations to find the solution of the gauge field that describes the vortex. Thus, we find the $a(r)$ solution, make an interpolation again and remember that the gauge field must satisfy the condition expressed in eq. (\ref{boundary}). With this, we present the behavior of the gauge field in fig. \ref{fig2}.

The magnetic field of the vortex is described by eq. (\ref{B}). So,  using the expression for magnetic field, we find the behavior this field associated to the generalized vortex.
\begin{figure}[ht!]
    \centering
    \includegraphics[scale=0.55]{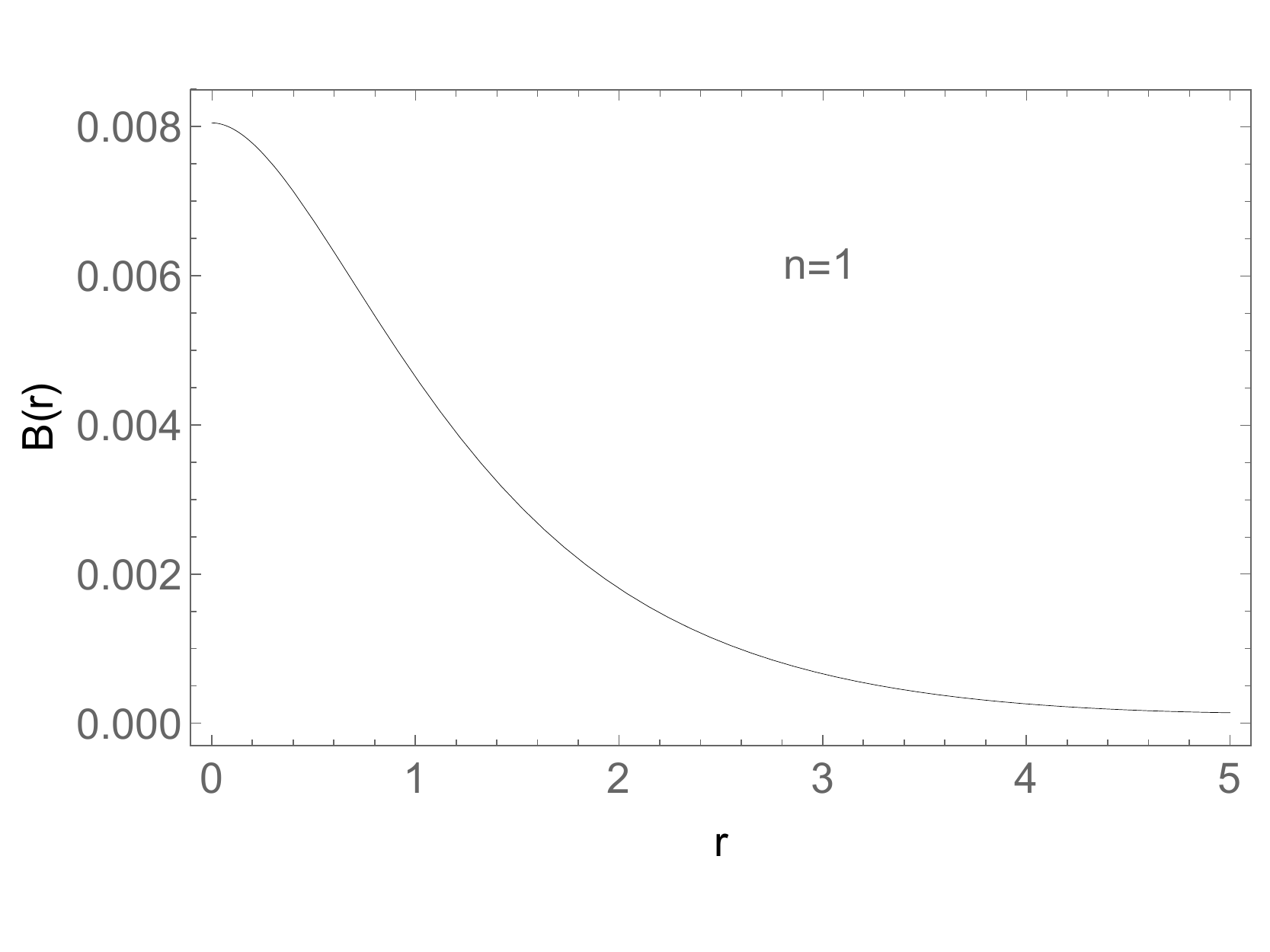}
    \caption{Magnetic fields of the vortex described by the solutions of BPS equation.}
    \label{fig3}
\end{figure}

\begin{figure}[ht!]
    \centering
    \includegraphics[scale=0.55]{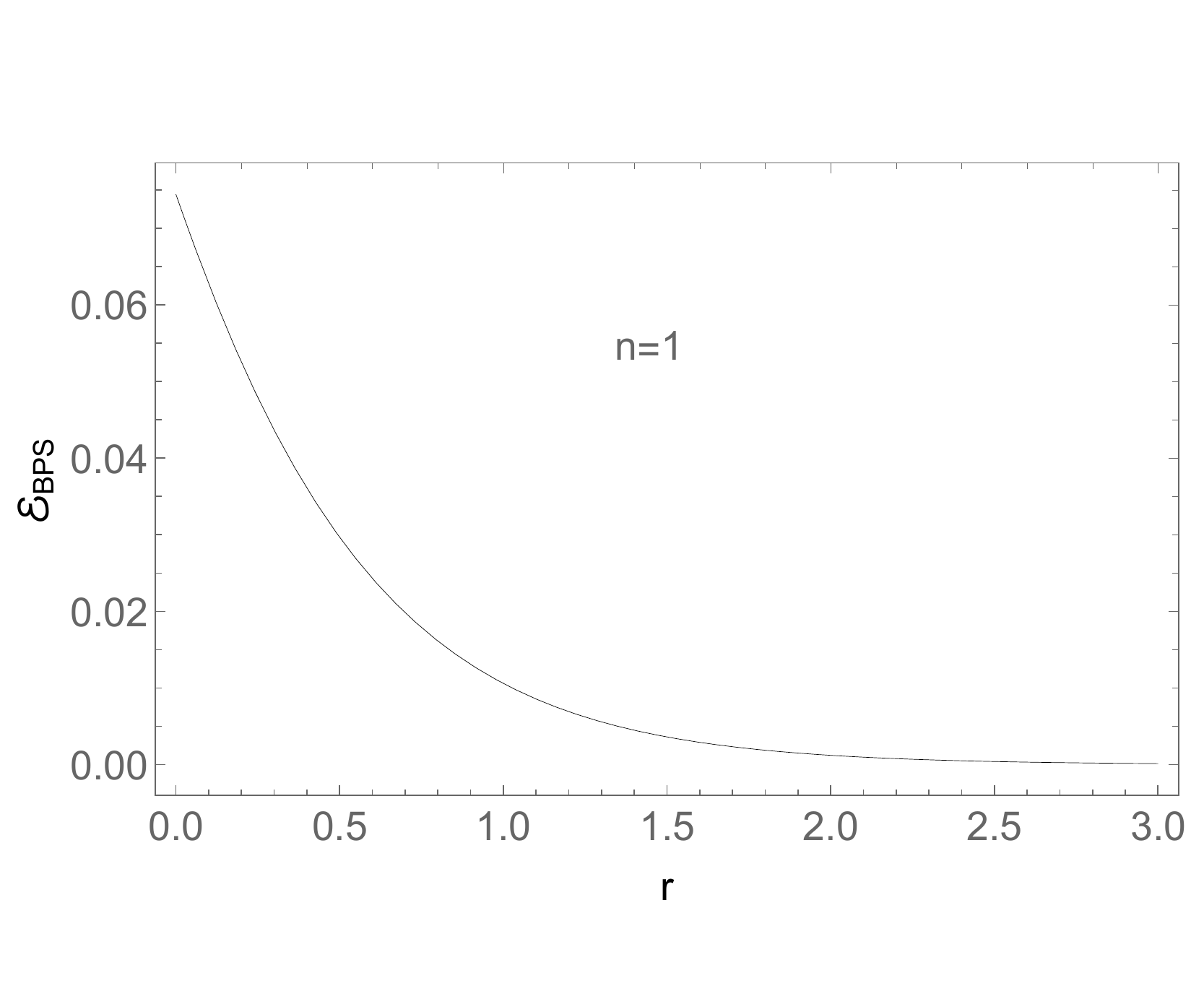}
    \caption{BPS energy density of the vortex.}
    \label{fig4}
\end{figure}

Considering eq. (\ref{EBPS}), we find the BPS energy density associated with the vortex, as shown in fig. \ref{fig4}.

We noticed that the vortices become more energetic in the around of $r=0$. This fact can be understood because the electrical permeability is of the logarithm type. It is interesting to mention that due to the kink-like topological solutions we found numerically, and the ansatz used, the vortices generated by the model will have a ring-like shape.

Finally, we conclude that as the magnetic field becomes more intense, we have more energetic vortices due to the increased magnetic flux. We observe also that both the BPS energy density and the magnetic field present similar shapes near the origin and differ only in their magnitude around $r\rightarrow 0$. 

\subsection{Electric permeability of the type $\rho f(|\phi|)^2\ln(m(|\phi|)^{2}/\mu^{2})$}

A more general model, discussed in the section II, appears when we consider the action being:
\begin{equation}
\label{action2}
    S=\int\, d^{3}x \bigg[-\frac{\rho f(|\phi|)^2}{4}\ln\bigg(\frac{m(|\phi|)^{2}}{\mu^{2}}\bigg)F_{\mu\nu}F^{\mu\nu}+|D_{\mu}\phi|^{2}-\mathcal{V}\bigg].
\end{equation}

The equations of motion associated with action (\ref{action2}) are:
\begin{align}
\label{m1}
    D_{\mu}D^{\mu}\phi+\frac{\phi}{2|\phi|}\bigg\{\frac{\rho f(|\phi|)}{2}\bigg[\frac{f(|\phi|)m_{|\phi|}}{m(|\phi|)}+f_{|\phi|}\ln\bigg(\frac{m(|\phi|)^2}{\mu^2}\bigg)\bigg]F_{\mu\nu}F^{\mu\nu}+\mathcal{V}_{|\phi|}\bigg\}=0,
\end{align}

\begin{align}
\label{m2}
    J^{\nu}=\partial_{\mu}\bigg[\rho f(|\phi|)^{2}\ln\bigg(\frac{m(|\phi|)^2}{\mu^2}\bigg)F^{\mu\nu}\bigg].
\end{align}
where we defined $f_{|\phi|}=df/d|\phi|$, $m_{|\phi|}=dm/d|\phi|$, and $\mathcal{V}_{|\phi|}=\partial\mathcal{V}/\partial |\phi|$.

The conserved current of the model again takes the form given by eq. (\ref{current2}). From eq. (\ref{m2}) we analyze the component $\nu=0$. Thus, we find again that the vortices must be electrically neutral. As expected, this occurs due to the fact that the electrical permeability function does not modify this property of Maxwell's vortices.

Now we find the energy-momentum tensor:
\begin{align}
  T_{\mu\nu}=\rho f(|\phi|)\ln\bigg(\frac{m(|\phi|)^{2}}{\mu^{2}}\bigg)F_{\mu\lambda}F^{\lambda}\, _{\nu}+\overline{D_{\mu}\phi}D_{\nu}\phi+\overline{D_{\nu}\phi}D_{\mu}\phi-g_{\mu\nu}\mathcal{L}.
\end{align}

Considering the ansatz for the variable fields and the conditions (\ref{boundary}), we observe again, that the magnetic field of the vortex will look like
\begin{align}
\label{MagneticField2}
    B=-(\partial_1 A_2-\partial_2 A_1)=-\frac{a'(r)}{er}.
\end{align}

From the above result, we conclude that the vortices of this model will also describe a quantized magnetic flux, i. e., $\Phi_{B}=2\pi n/e$.

Clearly, we are again interested in the study of the BPS vortices. In order to study these vortices we again use the energy-momentum tensor. In this case, the energy density can be written in the form:
\begin{align}
   \mathcal{E}=\frac{\rho f(g)^2 a'(r)^{2}}{e^{2}r^{2}}\ln\bigg(\frac{m(g)}{\mu}\bigg)+g'(r)^{2}+\frac{a(r)^{2}g(r)^{2}}{r^{2}}+\mathcal{V}(g)
\end{align}

Rearranging the energy density of the model, we will arrive at
\begin{align}\nonumber
\label{EN}
    \mathcal{E}=&\frac{\rho f(g)^2}{2}\ln\bigg(\frac{m(g)^{2}}{\mu^{2}}\bigg)\bigg[\frac{a'(r)}{er}\pm\frac{e(\xi^{2}-g(r)^{2})}{\rho f(g)^2\ln(m(g)^{2}/\mu^{2})}\bigg]^{2}+\bigg(g'(r)\mp\frac{a(r)g(r)}{r}\bigg)^{2}+\mathcal{V}\\
    & -\frac{e^{2}}{2}\frac{(\xi^{2}-g(r)^{2})^{2}}{\rho f(g)^2 \ln(m(g)^{2}/\mu^{2})}\mp\frac{1}{r}[a(r)(\xi^{2}-g(r)^{2})]'.
\end{align}

If we choose the potential to be
\begin{align}
    \mathcal{V}=\frac{e^{2}}{2}\frac{(\xi^{2}-g(r)^{2})^{2}}{\rho f(g)^2\ln(m(g)^{2}/\mu^{2})},
\end{align}
the expression for the energy density is simplified essentially. In this case, it is easy to find the BPS limit in the model, so that the Bogomol'nyi equations will look like:
\begin{align}
\label{B11}
    g'(r)=\pm\frac{a(r)g(r)}{r},
\end{align}
and 
 \begin{align}
 \label{B22}
     \frac{a'(r)}{er}=\mp\frac{e(\xi^{2}-g(r)^2)}{2\rho f(g)^2 \ln(m(g)/\mu)}.
 \end{align}
 
 As a consequence, the BPS energy density of the vortex is
 \begin{align}
 \label{EBPS4}
     \mathcal{E}_{BPS}=\frac{2\rho f(g)^{2}a'(r)^{2}}{e^{2}r^{2}}\ln\bigg(\frac{m(g)}{\mu}\bigg)+2g'(r)^{2}.
 \end{align}
 
From that point on, we must again turn our attention to the numerical study of topological structures described by the BPS equations (\ref{B11}) and (\ref{B22}).
 
 \subsubsection{The case: $f(g)=g(r)$ and $m(g)=hg(r)$.}
 
As in the previous cases, we return the BPS equations, i. e., eqs. (\ref{B11}) and (\ref{B22}), and using the interpolation method we investigate the field solutions. For simplicity, we will assume the study for the cases of $f(|\phi|)$ and $m(|\phi|)$ in the simplest cases, that is,
 \begin{align}
     f(g)=g(r) \hspace{1.5cm} \text{and} \hspace{1.5cm} m(g)=hg(r).
 \end{align}
 
\begin{figure}[ht!]
    \centering
    \includegraphics[scale=0.55]{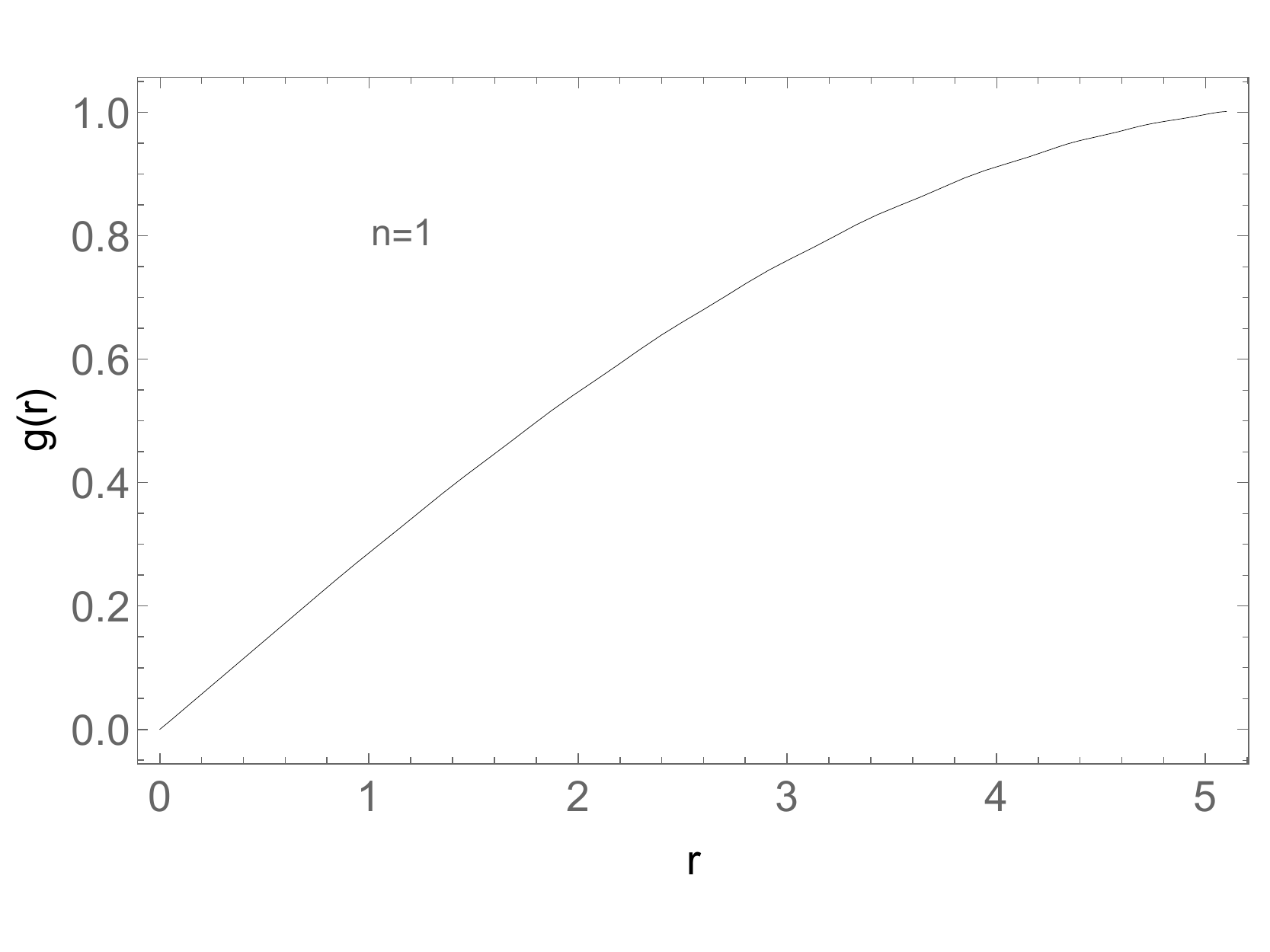}
    \caption{Solution of $g(r)$ corresponding to the complex scalar field when $\rho=10^{5}$, $e=\xi=1$, $h=1$ and $\mu=10^{-2}$.}
    \label{fig9}
\end{figure}

Assuming these functions and considering the boundary conditions (\ref{boundary}) of the model, we find the numerical solution expressed in fig. \ref{fig9}.

\begin{figure}[ht!]
    \centering
    \includegraphics[scale=0.55]{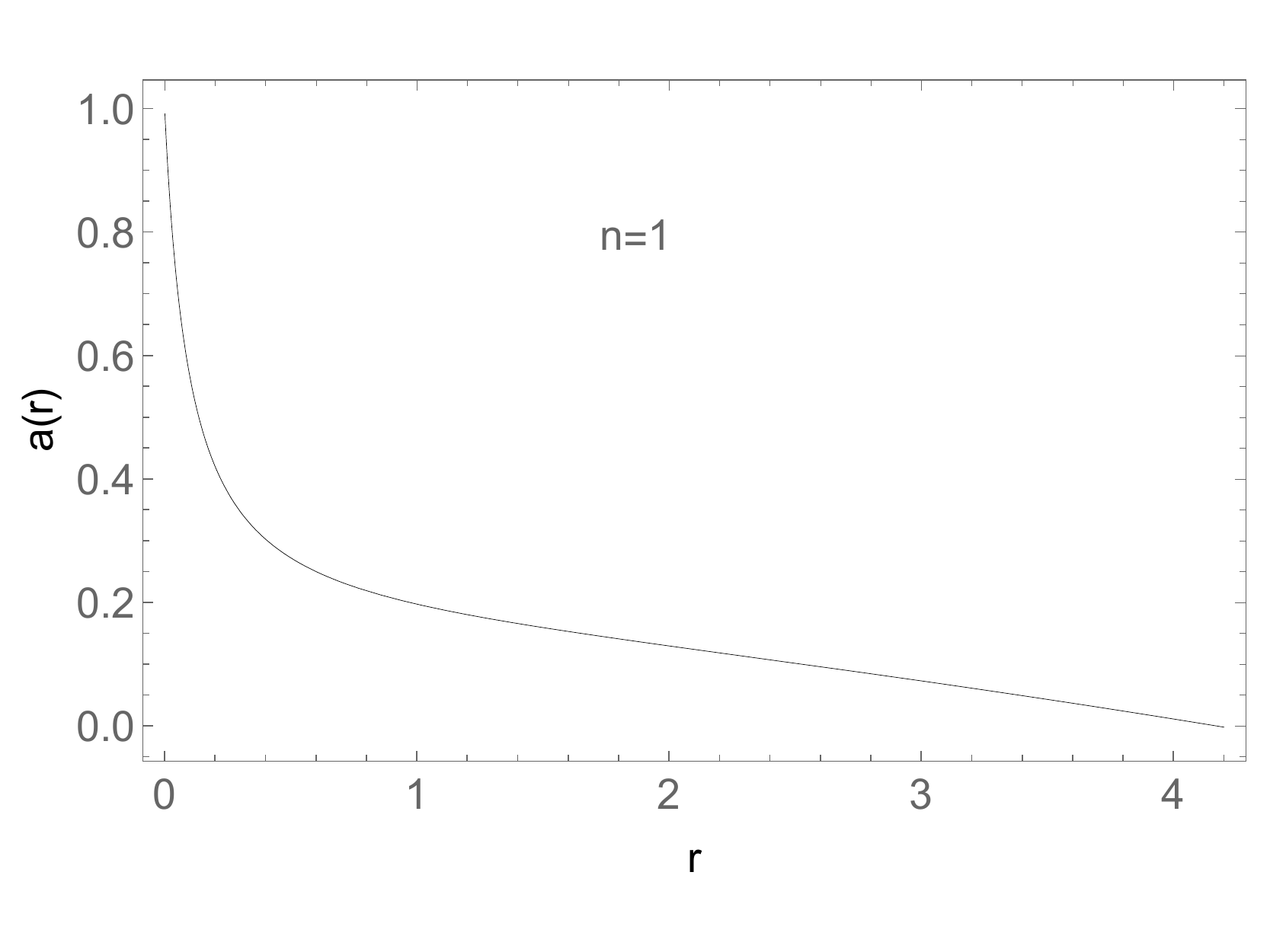}
    \caption{Gauge field $a(r)$ corresponding to the complex scalar field when $\rho=10^{5}$, $e=\xi=1$, $h=1$ and $\mu=10^{-2}$.}
    \label{fig10}
\end{figure}

With the solution for the variable field $g(r)$, we used the expression (\ref{B11}) to find the numerical solution of the gauge field. This solution is shown in fig. \ref{fig10}.

\begin{figure}[ht!]
    \centering
    \includegraphics[scale=0.55]{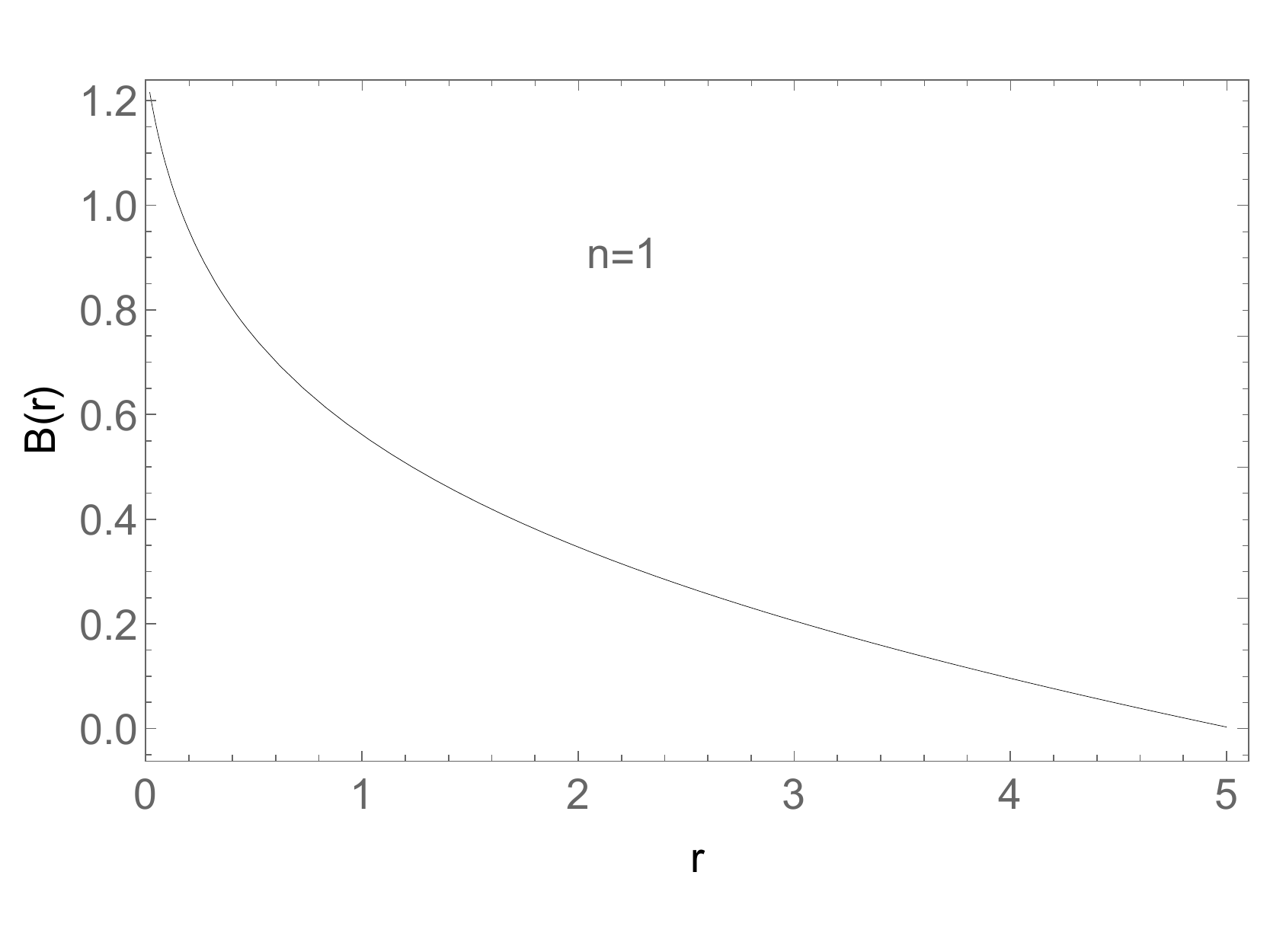}
    \caption{Magnetic fields of the vortex described by the solutions of BPS equation.}
    \label{fig11}
\end{figure}

Similar to the cases studied previously, we used eq. (\ref{MagneticField2}) to investigate the model's magnetic field. With this, we show the behavior of the magnetic field in fig. \ref{fig11}.

\begin{figure}[ht!]
    \centering
    \includegraphics[scale=0.55]{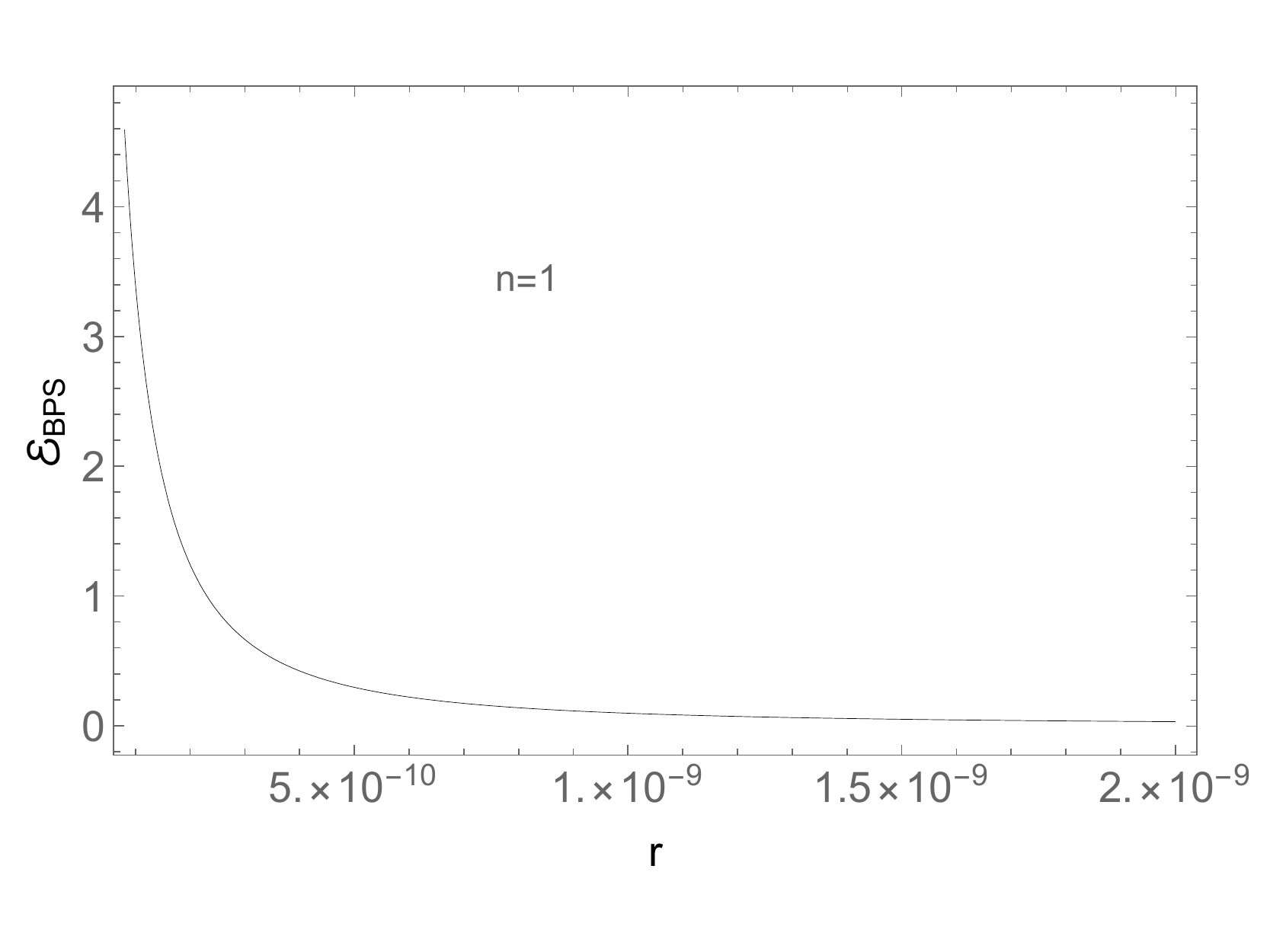}
    \caption{BPS energy density of the vortex.}
    \label{fig12}
\end{figure}

The energy density associated with the BPS vortices is described in the expression (\ref{EBPS4}). From this, we have the behavior of the model energy described in fig. \ref{fig12}.

Looking at the vortex structures presented above, we note that in the simplest case, i. e., $f(|\phi|)^{2}\propto |\phi|^2 $, it allows us to find topological structures of vortices when the amplitude of the electrical permeability function is of the order of $\rho=10^{5}$. These structures are much more energetic than the vortices studied in the previous cases ($f\rightarrow 1$). However, as we are working with models characterized by logarithmic permeability, we find topological vortices with extremely localized BPS energy around the origin. We note that the higher the order of the $f(|\phi|)$ polynomial, the more energetic is the vortex structure. As in previous cases, we observed that when $r\rightarrow 0$, the greater the magnetic flux of the vortex, and, in this way, the greater the vortex energy in that region.

\section{Discussions and conclusion}

In this work, we used a perturbative approach in scalar-vector theories to obtain some possible forms of the electrical permeability functions. In other words, we found the form that the model's generalization function should take, i. e., the logarithmic type function. Then, we studied a Maxwell-like theory generalized by a logarithmic scalar function multiplier. In our model, the gauge field is described by Maxwell term, but there is a coupling between the gauge field and the scalar field. Starting our investigation, we observed the existence of vortex structures in the model, with a gauge field generating an intense magnetic field in the neighborhood of $r=0$ and with quantized magnetic flux in all the cases studied. We found also that when the $f(|\phi|)$ function takes on polynomial forms, the vortices start to have a more intense magnetic flux making the energy of the vortex extremely located.

We demonstrated that the logarithmic vortices have an interesting property, the values of the vacuum state will only be reached at points far from the origin of the model. These characteristics are most visible when assume that the functions $f(|\phi|)$ and $m(|\phi|)$ are higher-order polynomials.

Clearly, as expected if $f(|\phi|)$ is assumed to be a constant function $(f=0)$, the topological vortices of the model no longer exist. However, it is possible to verify non-topological zero energy solutions. Obviously, these solitons would be trivial solitons of the model.

An extremely intriguing fact of the models with logarithmic generalizations studied so far, is the fact that the magnetic field and the energy density are extremely intense in $r\rightarrow 0$. This interesting result leads us to believe that this is due to the shape of the electrical permeability function.

Finally, it is observed that all the vortices studied in this work present quantized energy with the solutions being degenerate $\xi^{2}$ in a given topological sector.

During all work, we investigated BPS vortex structure with logarithmic generalization. We concluded that the electrical permeability function must have a logarithmic form. With this, we observed that the permeability directly influences the dynamics of the gauge field and complex scalar field that describe Maxwell's vortices. We conclude that the vortex with logarithmic generalization have well-located and intense BPS energy in the region where the gauge field is maximum, this is due to the high magnetic flux in that region. Finally, we can note that all vortices in the model are degenerated by a factor of $\xi^2$. 

The further continuation of this study could consist in consideration of Lorentz-breaking generalizations of these models. We expect to perform this study in a forthcoming paper.


\section*{Acknowledgments}
The authors thank the Conselho Nacional de Desenvolvimento CientÃ­fico e Tecnol\'{o}gico (CNPq), grant n$\textsuperscript{\underline{\scriptsize o}}$ 308638/2015-8 (CASA), grant n$\textsuperscript{\underline{\scriptsize o}}$ 301562/2019-9 (A. Yu. Petrov), and Coordena\c{c}ao de Aperfei\c{c}oamento do Pessoal de N\'{\i}vel Superior (CAPES), for financial support. Authors are grateful to P. Porfirio for important discussions.


\end{document}